\documentclass[reprint, superscriptaddress,nofootinbib
nofootinbib,
nobibnotes,
amsmath,amssymb,
aps,
pra,
floatfix,
]{revtex4-2}

\usepackage{graphicx}
\usepackage{dcolumn}
\usepackage{bm}
\usepackage{color}
\usepackage{ulem}
\usepackage{comment}
\usepackage[colorlinks=true,urlcolor=blue,linkcolor=blue,citecolor=blue]{hyperref}

\newcommand{\C}{\mathbb{C}} 
\newcommand{\R}{\mathbb{R}} 
\newcommand{\N}{\mathbb{N}} 
\renewcommand{\Re}{\operatorname{Re}}
\renewcommand{\Im}{\operatorname{Im}}
\newcommand{\iu}{\mathrm{i}}
\newcommand{\eu}{\mathrm{e}}
\newcommand{\eps}{\varepsilon}
\newcommand{\de}{\mathrm{d}}

\newcommand{\zetanew}{\zeta_\text{new}}

\DeclareMathOperator{\argmax}{arg\,max}

\newcommand\un[1]{\underline{#1}}

\begin{document}


\title{Enhanced Soliton Stability in Bi-directionally Coupled Laser-Microresonator Systems}

\author{L.~Bengel}
\thanks{These authors contributed equally: L.~Bengel, H.~Peng}
\affiliation{Institute for Analysis (IANA), Karlsruhe Institute of Technology, 76131 Karlsruhe, Germany}

\author{H.~Peng}
\thanks{These authors contributed equally: L.~Bengel, H.~Peng}
\affiliation{Institute of Photonics and Quantum Electronics (IPQ), Karlsruhe Institute of Technology, 76131 Karlsruhe, Germany}

\author{B.~de Rijk} 
\affiliation{Institute for Analysis (IANA), Karlsruhe Institute of Technology, 76131 Karlsruhe, Germany}


\author{C.~Koos}
\affiliation{Institute of Photonics and Quantum Electronics (IPQ), Karlsruhe Institute of Technology, 76131 Karlsruhe, Germany}

\author{W.~Reichel}
\email[]{wolfgang.reichel@kit.edu}
\affiliation{Institute for Analysis (IANA), Karlsruhe Institute of Technology, 76131 Karlsruhe, Germany}

\date{\today}

\begin{abstract} 
We investigate a bi-directionally coupled system consisting of a Kerr-nonlinear microresonator and a continuous-wave single-mode semiconductor laser. Inside the resonator, a forward-propagating and a backscattered field interact nonlinearly, while a fraction of the backscattered field is fed back into the laser cavity. We show in this paper that the interaction of the laser with the feedback opens up new ways of stabilizing $1$-solitons. Using numerical bifurcation analysis, we systematically identify existence ranges of time-harmonic 1-soliton states in the anomalous dispersion regime. We demonstrate that, in contrast to the uni-directional configuration, the bi-directional coupling introduces a dynamic self-correcting response of the laser frequency that stabilizes $1$-solitons. These enhanced stability properties of $1$-solitons thus enable robust and self-started frequency-comb generation, consistent with the existing experimental observations.
\end{abstract}

\maketitle

\section{Introduction}

Chip-scale Kerr soliton comb generators using integrated optical microresonators have emerged as novel light sources that offer vast application potential. Providing tens or even hundreds of narrow-linewidth phase-locked tones with equidistant spacings of tens or hundreds of gigahertz (GHz), the devices are, e.g., particularly promising for massively parallel optical communications~\cite{Marin-Palomo:20}, low-noise microwave generation~\cite{ji2025dispersive,jin2025microresonator,sun2024integrated,Zhao2024}, or ultra-broadband photonic-electronic signal processing~\cite{Drayss:23,fang2025320,11190945,11262997}. Recently, tremendous progress has been made regarding ultra-low-loss optical waveguides, enabling Kerr-nonlinear microresonators with extremely high-$Q$ factors of the order $10^7$ or even $10^8$~\cite{liu2021high,ji2025deterministic,Chen2026}. This results in milliwatt-level pump thresholds for Kerr comb generation~\cite{Shen2020Integrated}, thereby allowing for pumping by III-V-based semiconductor laser diodes with moderate output power. A particularly promising approach to compact robust Kerr comb sources relies on directly coupling the pump laser diode to the Kerr-nonlinear microresonator without any intermediate optical isolator, leveraging resonantly enhanced backscattering from the cavity to precisely lock the laser emission wavelength to the pumped resonance~\cite{Shen2020Integrated,Xiang2021Laser,Lihachev2022Platicon}. This approach has been demonstrated both experimentally and theoretically to offer fully automatic soliton-comb formation by simply switching on the laser pump~\cite{Shen2020Integrated,Raja2019Electrically,Xiang2021Laser,Lihachev2022Platicon,Voloshin2021Dynamics}, thereby obviating complex electronic control systems and allowing for comb sources with compact footprint that can be integrated into standard industrial packages. However, while a series of experimental demonstrations of self-injection-locked Kerr comb generators has been reported over the previous years~\cite{Shen2020Integrated,Raja2019Electrically,Xiang2021Laser,Lihachev2022Platicon}, theoretical analysis and modeling was so far limited to conventional time-integration techniques, that can describe the field evolution in a specific comb-source implementations, but that cannot give a full picture of the stability and the properties of such comb states over a wider parameter range.

In this paper we address this gap and develop a framework that allows to investigate soliton states of the bi-directionally coupled system using numerical bifurcation analysis in the anomalous dispersion regime. In contrast to time-integration methods, which were used in previous studies~\cite{Shen2020Integrated,Xiang2021Laser,Lihachev2022Platicon}, our numerical bifurcation approach allows for a systematic exploration of soliton states for a large range of technically relevant parameter values. Our numerical analysis shows that, unlike its one-directionally coupled counterpart, the bi-directionally coupled laser-microresonator system can self-correct perturbations of the laser frequency such that stable $1$-solitons persist. We supplement our study by a sensitivity analysis of the $1$-soliton states with respect to parameter variations. 


The paper is structured as follows. In Section~\ref{sec:equations} we formulate the mathematical model for the bi-directionally coupled system. In Section~\ref{sec:comparison} we present numerical simulations and compare our findings with those obtained in the standard uni-directionally coupled laser-microresonator model. In Appendix~\ref{app:numerical_bifurcation} we give details on the numerical bifurcation analysis as well as the algorithm for the parameter sensitivity analysis. Appendix~\ref{app:normalization} is devoted to the derivation of the normalized form of the bi-directionally coupled system from the system in physical quantities.

\section{Coupled laser-resonator model equations} \label{sec:equations}
\begin{figure}
    \centering
    \includegraphics[width=\columnwidth]{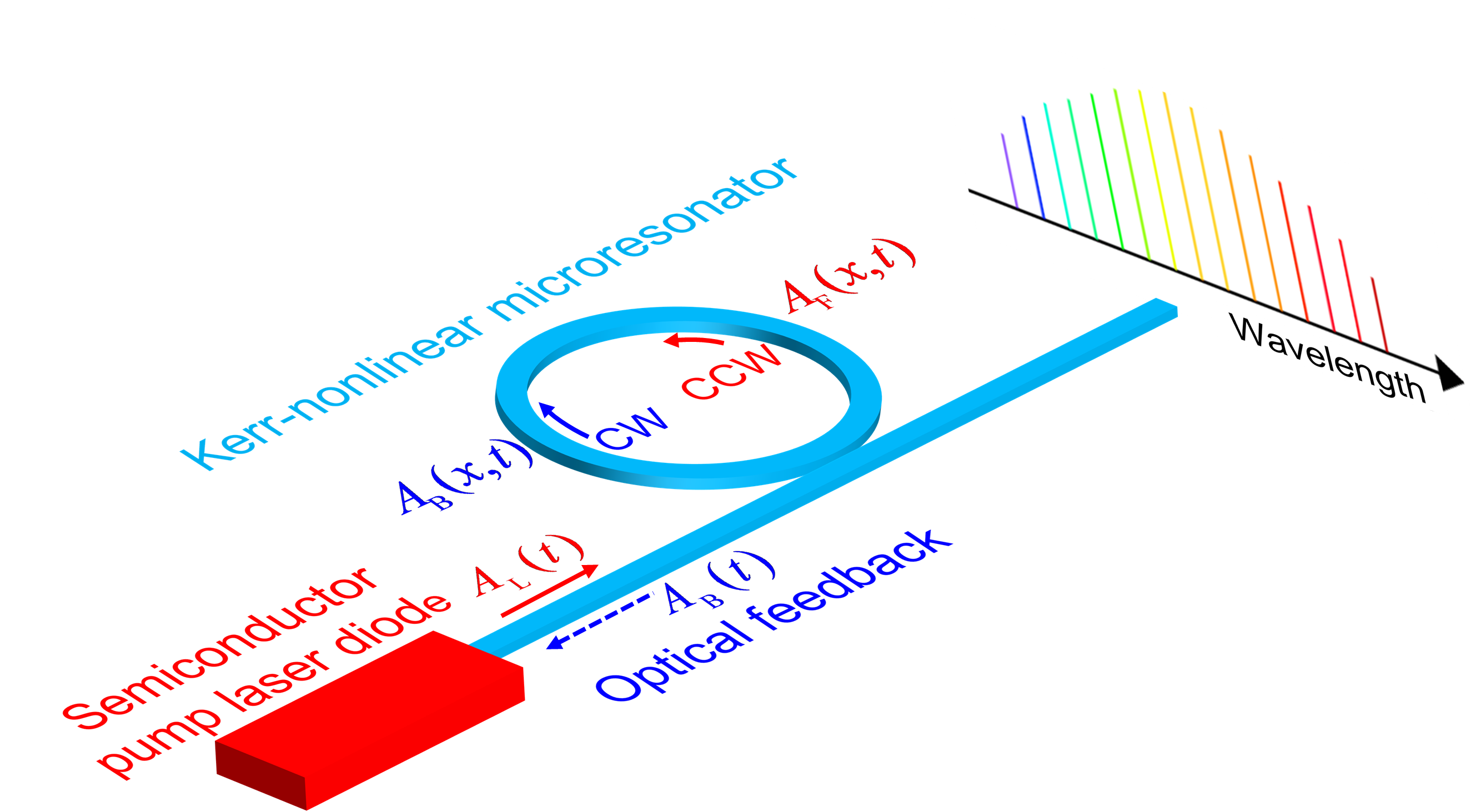}
    \caption{Schematic diagram depicting the bi-directional coupling of the semiconductor laser with a Kerr microresonator. The light emitted from the semiconductor laser is injected into the Kerr microresonator in counter-clockwise (CCW) direction. Due to surface and internal inhomogeneities, the CCW field is backscattered clockwise (CW) and injected back into the semiconductor laser.}
    \label{fig:cw_ccw}
\end{figure}

Direct pumping of a Kerr-nonlinear microresonator with a semiconductor laser diode can be considered as a bidirectionally coupled system, with resonantly enhanced Rayleigh backscattering within the microresonator feeding back into the laser cavity, see the conceptual diagram of the coupled system in Figure~\ref{fig:cw_ccw}. The light emitted from the pump laser is described by the carrier number $N(t)$ inside the cavity and the laser field $A_\text{L}(t)$ which is coupled into the ring-shaped Kerr microresonator. A counter-clockwise (CCW) propagating forward field $A_\text{F}(x,t)$ and a clockwise (CW) propagating backscattered field $A_\text{B}(x,t)$ are generated, where $x \in [0,2\pi]$ is the angular position and $t \geq 0$ is the time. Inside the resonator the fields $A_\text{F}$ and $A_\text{B}$ interact with each other and a part of $A_\text{B}$ is coupled out of the resonator back into the laser where it interacts with the laser field $A_\text{L}$.
On a fundamental level, the Kerr-nonlinear interaction in a high-$Q$ microresonator pumped by a strong cw-laser without backcoupling is described by the Lugiato-Lefever equation (LLE)~\cite{Lugiato1987Spatial,HAELTERMAN1992401}:
\begin{equation} \label{eq:lle}
\dot a(x,t) = \Bigl[ - 1 + \iu d_2 \frac{\partial^2}{\partial x^2} + \iu |a(x,t)|^2\Bigr] a(x,t) - \kappa_\text{ext} f \eu^{\iu( \zeta t+ \phi)}
\end{equation}
in normalized time $t$ and angular position $x\in [0,2\pi]$  (a superscripted dot denotes the time derivative). Here $a$ is $2\pi$-periodic in $x$ and describes the optical field in the microresonator driven by the invariant laser field $a_\text{L}(t) = f \eu^{\iu\zeta t}$ where the detuning $\zeta=\omega_\text{m}-\omega_0$ is static and is given by the mismatch between the frequency $\omega_\text{m}$ of the pumped resonator mode and the frequency $\omega_0$ of the laser cavity mode. Dispersion $d_2$ and coupling $\kappa_\text{ext}$ are as in Table~\ref{constants_table} and $f$ represents the strength of the input laser.

\medskip

The set-up in this paper differs from the one described by the LLE \eqref{eq:lle} in the sense that the backscattered field interacts with the laser. This leads to a bi-directionally coupled system, whose dynamics is governed by the equations~\cite{Xiang2021Laser,Lihachev2022Platicon}:
\begin{align} \label{eq:coupled_system}
    \begin{split}
    \dot n(t) =& \iota - \gamma n(t) - g(|a_\text{L}(t)|^2) (n(t)-1) |a_\text{L}(t)|^2, \\
    \dot a_\text{L}(t) =& 
    \Bigl[\frac{1-\iu\alpha_\text{H}}{2} \left(g(|a_\text{L}(t)|^2) n_0 (n(t)-1)- \kappa_\text{d}\right)\\
    & + \iu (\omega_\text{m}-\omega_0)\Bigr]a_\text{L}(t)- \kappa_\text{ext} a_\text{B}(t)\eu^{\iu \phi},\\
    \dot a_\text{F}(x,t) =& \Bigl[ - 1 + \iu d_2 \frac{\partial^2}{\partial x^2} + \iu (|a_\text{F}(x,t)|^2 + 2 |a_\text{B}(t)|^2)\Bigr] \\
    & \times a_\text{F}(x,t) + \iu \kappa_\text{sc}a_\text{B}(t) - \kappa_\text{ext} a_\text{L}(t) \eu^{\iu \phi}, \\
    \dot a_\text{B}(t) =& \Biggl[ -1 + \iu \bigl( |a_\text{B}(t)|^2 + \frac{1}{\pi} \int_0^{2\pi} |a_\text{F}(x,t)|^2 \de x \bigr)\Biggr] \\
    & \times a_\text{B}(t) + \iu \overline{\kappa}_\text{sc} \frac{1}{2\pi} \int_0^{2\pi} a_\text{F}(x,t) \de x,
    \end{split}
\end{align}
where $a_\text{F}$, $a_\text{B}$, and $a_\text{L}$ represent the normalized forward, backward, and laser fields, respectively, and $n$ denotes the normalized carrier number. For simplicity, the backscattered field $a_\text{B}$ is only represented by its central mode $a_\text{B}(t) := \frac{1}{2\pi}\int_0^{2\pi} a_\text{B}(x,t)\de x$ which is independent of the angular variable $x$. The derivation of the system~\eqref{eq:coupled_system} for the normalized unknowns $a_\text{F}, a_\text{L}, a_\text{B}, n$ from the system~\eqref{eq:coupled_system_Physical} for the physical quantities $A_\text{F}, A_\text{B}, A_\text{L}$, $N$ is detailed in Appendix~\ref{app:normalization}. Again, $a_\text{F}(x,t)$ is $2\pi$-periodic in $x$. The first two ordinary differential equations (ODEs) describe the laser dynamics for the carrier number $n$ and laser field $a_\text{L}$ with an additional coupling to the backscattered field $a_\text{B}$. Due to saturation, the normalized gain $g=g(|a_\text{L}|^2)$ of the laser is modeled by a decreasing function of the power $p=|a_\text{L}|^2$ of the laser field and takes the following explicit form 
\begin{align*}
    g(p) = \frac{\mu^2\sigma}{g_0 + \eps p}.
\end{align*}
The third and fourth equation describe the microresonator dynamics by a coupled system of a partial differential equation (PDE) for the forward field $a_\text{F}$ and an ODE for the backscattered field $a_\text{B}$ with the laser field $a_\text{L}$ as a source term for the forward field. 
All normalized constants are explained in Table~\ref{constants_table}. The values of the physical constants and their relation to the normalized constants can be found in Table~\ref{tab:constants} in Appendix~\ref{app:normalization}.

\begin{table}
\begin{tabular}{|c|c|l|} \hline
quantity & normal. & physical meaning \\ 
& quantity & \\ \hline
$I/e$  & $\iota$ &$\begin{array}{cc}\text{bias current of laser/} \\ \text{elementary electronic charge} 
\end{array}$\\
$\Gamma$ & $\gamma$ & carrier recombination rate \\
$N_0$ & $n_0$ & carrier number at transparency \\
$\mu$ & $-$ & optical mode confinement factor \\
$\alpha_\text{H}$ & $-$ & linewidth enhancement factor \\
$K_\text{d}$ & $\kappa_\text{d}$ & photon decay rate of laser cavity  \\
$\Omega_0$ & $\omega_0$  & laser cavity mode frequency \\
$K_\text{ext}$ & $\kappa_\text{ext}$ & external coupling \\
$K$ & $-$ & photon decay rate of resonator \\
$\Omega_\text{m}$ & $\omega_\text{m}$ &frequency of resonator mode \\
$D_2$ & $d_2$ & group velocity dispersion \\
$g_0$ & $-$ & single photon induced Kerr \\
& & frequency shift \\
$K_\text{sc}$ & $\kappa_\text{sc}$ & $\C$-valued coupling between forward\\
& & and backward modes  \\ 
$\phi$ & $-$ & accumulative phase between laser \\
& & and resonator \\
$\sigma$ & $-$ & small-signal gain\\
$\epsilon$ & $\eps$ & saturation coefficient\\
\hline
\end{tabular}
\caption{Description of the physical meaning of the constants and their normalized counterparts. A dash indicates that there is no normalized version of the corresponding quantity in the equations.}
\label{constants_table}
\end{table}

\section{Results and comparison between uni-directionally and bi-directionally coupled system} \label{sec:comparison}
Here, we present the results of our numerical bifurcation analysis for the bi-directionally coupled model \eqref{eq:coupled_system} and compare them with those for the uni-directionally coupled system \eqref{eq:lle}. An important part of the model is the normalized instantaneous laser frequency $\omega_\text{L}(t)$, which is defined by $\omega_\text{L}(t)-\omega_\text{m} := -\frac{d}{dt} \arg(a_\text{L}(t))$. In the standard LLE-model \eqref{eq:lle} we have that $\omega_\text{L}(t)=\omega_0$ is fixed whereas in the extended model \eqref{eq:coupled_system} it is in general time-dependent and part of the unknowns.

\medskip

Typical observations in the standard LLE~\eqref{eq:lle} are the following:
\begin{itemize}
\item[(i)] Stable $1$-soliton solutions of the form $a(x,t)= \un{a}^0(x)\eu^{\iu(\zeta t+\phi)}$, $\zeta = \omega_\text{m}-\omega_0$ can be determined by a bifurcation analysis,~\cite{Gaertner2019Bandwidth,Parra-Rivas2016Dark,Parra-Rivas2018BifurcationPeriodic,Parra-Rivas2018BifurcationLocalized}.
\item[(ii)] Stability means that, while keeping the laser in its prescribed invariant state, small perturbations of the resonator field $a$ relax to a translate of the $1$-soliton on an exponentially fast time scale~\cite{Bengel2024Stability,BengelSolitonBased,Godey2014Stability,Stanislavova2018Asymptotic}.
\end{itemize} 
For the extended system \eqref{eq:coupled_system}, instead, we observe the following:
\begin{itemize}
\item[(iii)]  A bifurcation analysis reveals stable time-harmonic $1$-soliton solutions 
\begin{align}
\begin{split}\label{eq:soliton_state}
a_\text{F}(x,t) &= \un{a}_\text{F}^0(x)\eu^{\iu\zetanew t}, \qquad n(t)  = n^0\\
a_\text{L}(t) &= \un{a}_\text{L}^0 \eu^{\iu\zetanew t}, \qquad a_\text{B}(t) = \un{a}_\text{B}^0 \eu^{\iu \zetanew t}.
\end{split}
\end{align} 
These $1$-solitons bifurcate from spatially constant time-harmonic states with $\zeta=\omega_\text{m}-\omega_0$ as the bifurcation parameter. They are qualitatively very similar to the ones found in \eqref{eq:lle} and the laser is still operating in a cw-mode with constant detuning $\zetanew=\omega_\text{m}-\omega_\text{L}=\zeta-\Delta\zeta$ where $\Delta\zeta = \frac{\kappa_\text{ext}}{\un{a}_\text{L}^0} \Im(\un{a}_\text{B}^0\eu^{\iu\phi}(1-\iu \alpha_\text{H}))$.
\end{itemize} 

\begin{figure*}
    \centering
    \includegraphics[width=.41\textwidth]{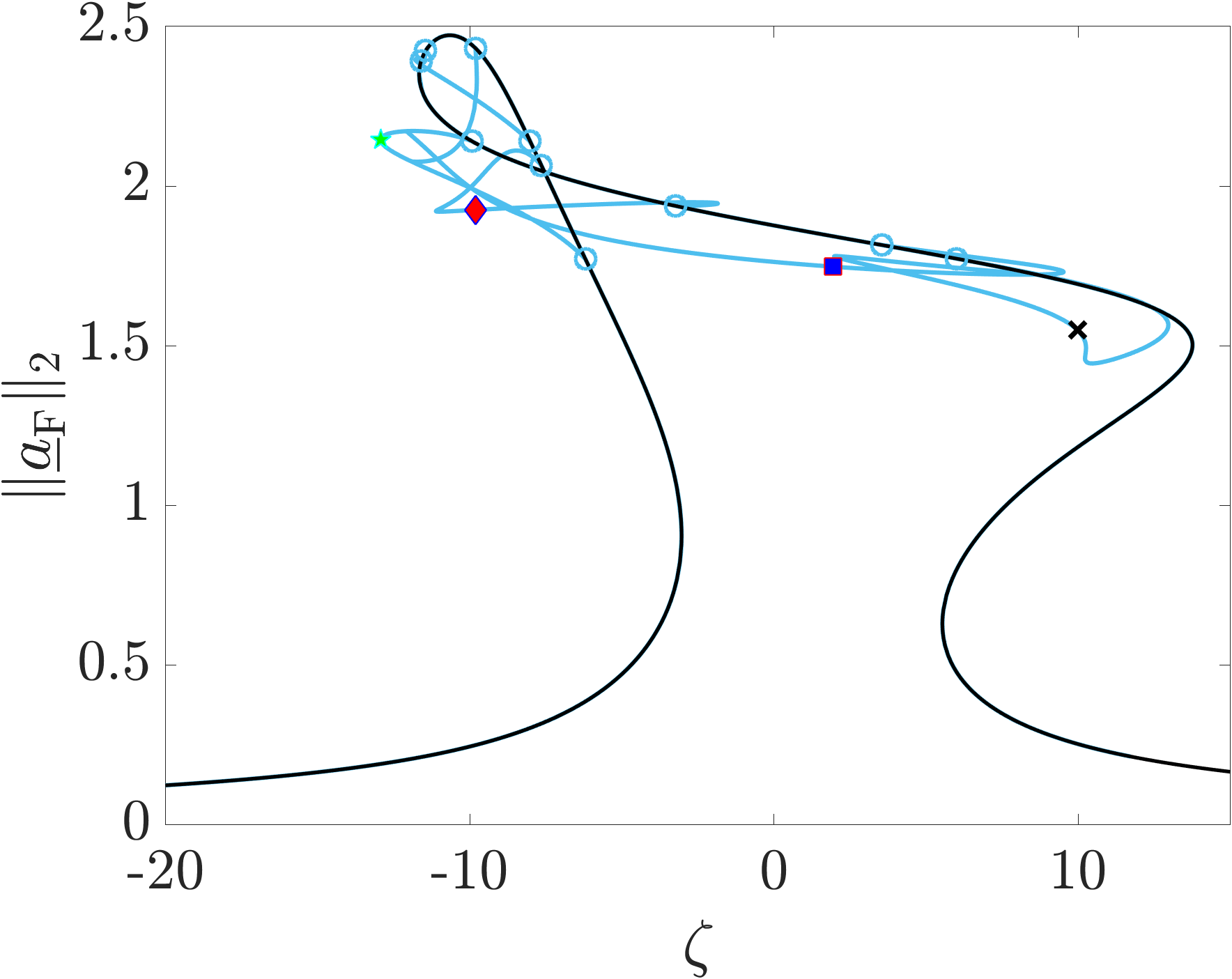} \hspace{4em}
    \includegraphics[width=.5\textwidth]{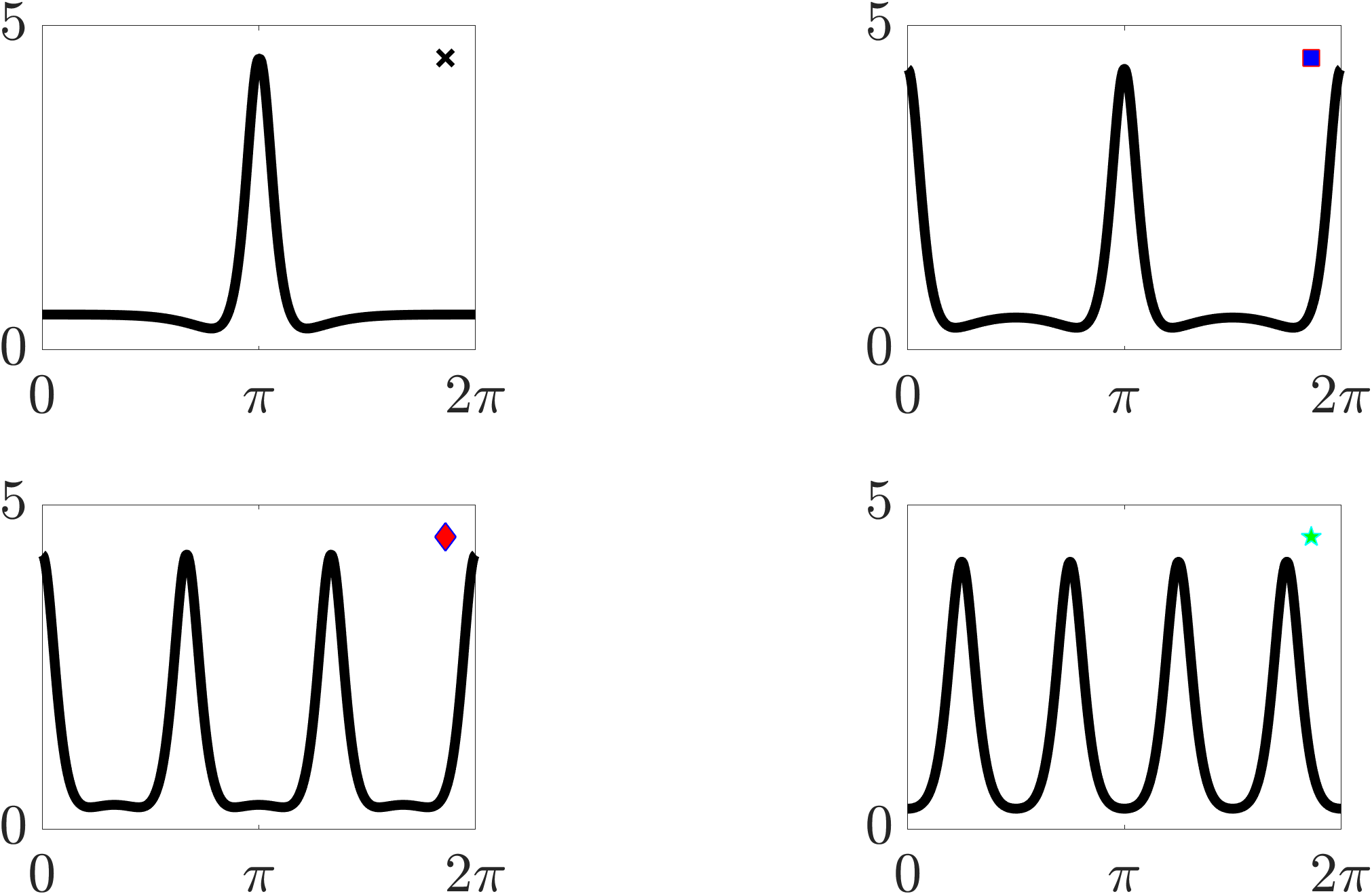}
    \caption{Left: Bifurcation diagram of stationary solutions to~\eqref{eq:coupled_system}, with physical constants set as in Table~\ref{tab:constants} and $\phi = 0.64\times\pi\approx 2.0106,\kappa_\text{ext}=6.4550$. As bifurcation parameter we choose the frequency difference $\zeta$. Right: Plots of the forward field component $\un{a}_\text{F}$ at the markers, which are localized states and the $y$-axis shows $|\un{a}_\text{F}|^2$.}
    \label{fig:bif_run1}
\end{figure*}

We illustrate this bifurcation phenomenon in Figure~\ref{fig:bif_run1} by a numerically computed bifurcation diagram together with examples of the localized steady states. 
The $y$-axis plots the power 
$$
    \|\un{a}_\text{F}\|_2 = \left(\int_0^{2\pi} |\un{a}_\text{F}(x)|^2\de x\right)^{1/2}
$$
of the forward-field component of the solution in the bifurcation diagram. The black line in the left panel of Figure~\ref{fig:bif_run1} indicates the curve of homogeneous constants solutions with the blue circles corresponding to the bifurcation points (BP). Continuation at the BPs (blue curves) then gives rise to many localized states (1-,~2-, 3-, and 4-soliton solutions) which are depicted in the panels on the right.  Most importantly, we find that the 1-soliton solutions lie on the bifurcation branch which bifurcates from the BP with the largest value of $\zeta$. As we follow the curve of constant solutions starting from $\zeta = -20$, this corresponds to the last BP in the diagram. The observation, that the 1-solitons are found on the last bifurcation branch is consistent with the theory developed for the standard LLE without backward laser coupling~\cite{Gaertner2019Bandwidth}. 

\begin{figure*}[t]
    \centering
    \includegraphics[width=.41\textwidth]{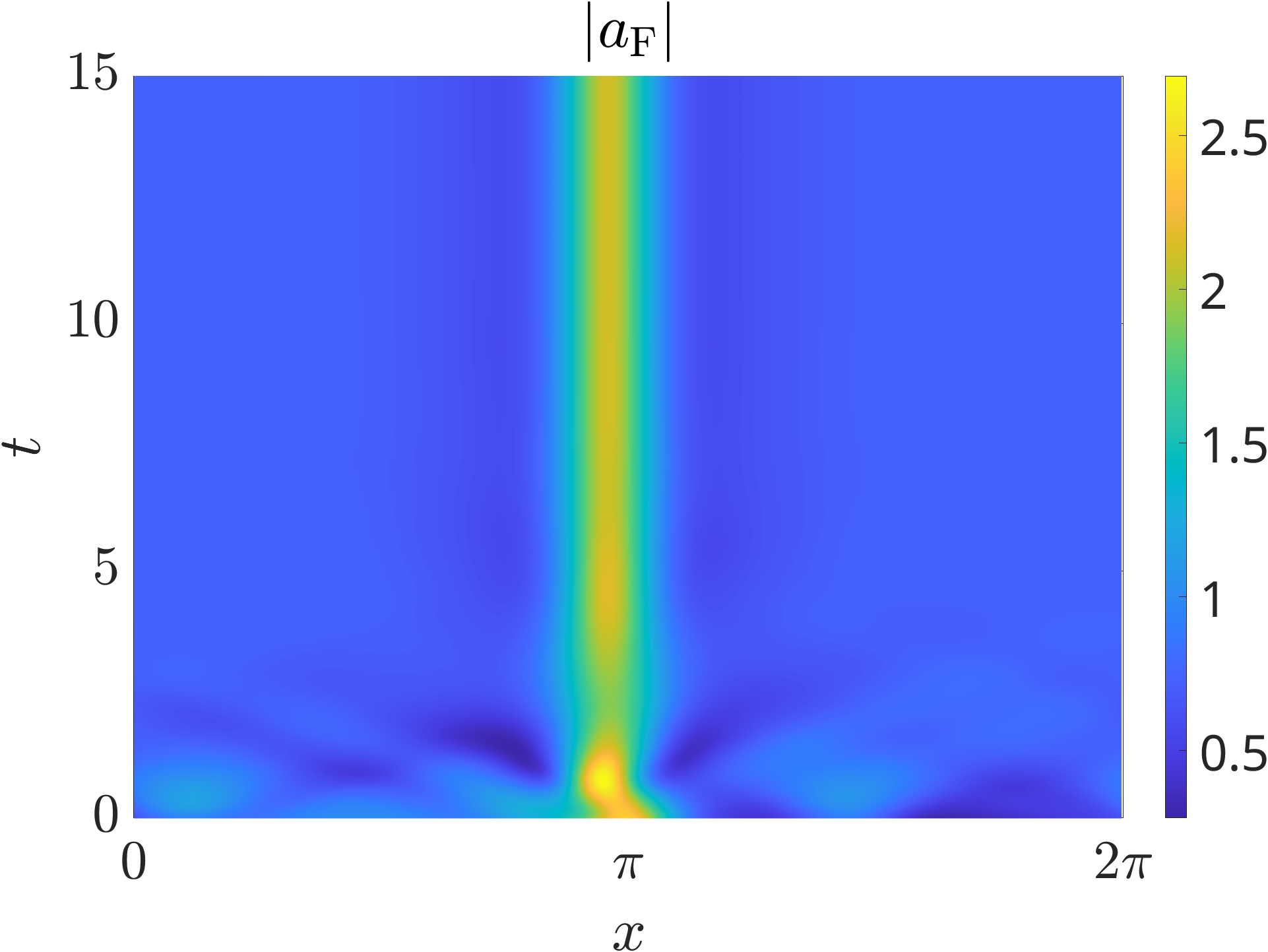}\hspace{4em}
    \includegraphics[width=.49\textwidth]{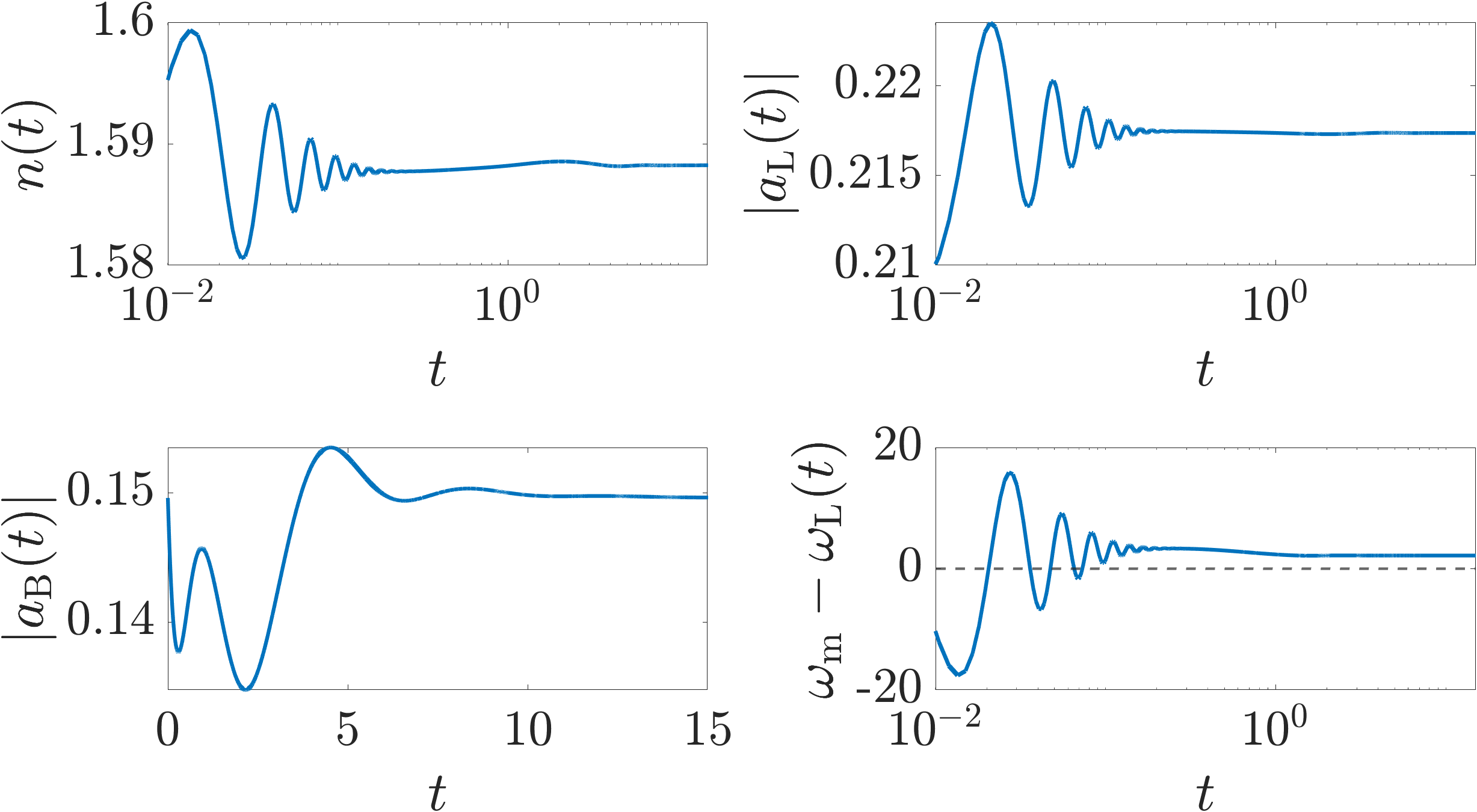}
    \caption{Time integration starting from a small perturbation of the stable 1-soliton solution with label \textbf{x} in Figure~\ref{fig:bif_run1}. Left: Space-time plot of the forward field $a_\text{F}(x,t)$, where the color code indicates the magnitude $|a_\text{F}(x,t)|$. Right: Simulations of the carrier density $n(t)$ (top left), magnitude of the laser field $|a_\text{L}(t)|$ (top right), the backward field $|a_\text{B}(t)|$ (bottom left), and the instantaneous laser frequency $\omega_\text{m}-\omega_\text{L}(t)$ (bottom right).}
    \label{fig:time-integration}
\end{figure*}

An important difference to \eqref{eq:lle} is the stability behavior of the solitons.
\begin{itemize}
\item[(iv)] We stress that stability of 1-solitons as solutions to the extended model (2) implies robustness against perturbations in all four unknowns $n, a_\text{L}, a_\text{F}$ and $ a_\text{B}$. In particular, applying such perturbations to the soliton solutions~\eqref{eq:soliton_state}, we find solutions $a_\text{F}(x,t)$ with the property that
\begin{align}
    \omega_\text{L}(t) &\to \omega_0+\Delta\zeta, \label{eq:laser_frequ_convergence} \\
    a_\text{F}(x,t) & \to \un{a}_\text{F}^0(x+x_0)\eu^{\iu(\zetanew t+\phi_0)}
\end{align}
with $\Delta\zeta = \frac{\kappa_\text{ext}}{\un{a}_\text{L}^0} \Im(\un{a}_\text{B}^0\eu^{\iu\phi}(1-\iu \alpha_\text{H}))$ exponentially fast as $t \to \infty$. Likewise 
\begin{align*}
    a_\text{L}(t) &\to \un{a}_\text{L}^0\eu^{\iu(\zetanew t+\phi_0)}, \quad n(t) \to n^0, \\
    a_\text{B}(t) &\to \un{a}_\text{B}^0\eu^{\iu(\zetanew t+\phi_0)}  & 
\end{align*}
at an exponential rate. This means that due to the backcoupling the laser automatically adjusts and self-corrects towards a time-harmonic cw-mode with constant detuning. We emphasize that this self-correction of the laser found in the extended system~\eqref{eq:coupled_system} cannot be observed in \eqref{eq:lle} due to the static form of the laser field. 
\item[(v)] Here $\phi_0 = \int_0^\infty (\omega_\text{L}(s)-\omega_\text{L})\,ds$ is an accumulated phase and $x_0$ is a translational displacement of the asymptotic steady state $a_\text{F}^0$ induced by the perturbation, see Appendix~\ref{app:numerical_bifurcation} for more details. The translational shift is visible in the left panel of Figure~\ref{fig:time-integration}. Its value is fully determined by the initial perturbation.
\end{itemize}

As described in (iv), the soliton solutions to~\eqref{eq:coupled_system} have enhanced dynamical stability features as opposed to \eqref{eq:lle}. To illustrate these features, we use numerical time integration. To this end, we employ a Lie-Trotter splitting to approximate the dynamics with the perturbed soliton state \textbf{x} of Figure~\ref{fig:bif_run1} as an initial condition. The simulations in Figure~\ref{fig:time-integration} confirms that a perturbation of a stable 1-soliton solution causes all four unknowns to relax against a steady state. In particular, as $t\to\infty$ we see that $r(t):=|a_\text{L}(t)|$ and $\un{a}_\text{B}(t)$ converge to $r^\textbf{x}$ (magnitude of laser field) and $\un{a}_\text{B}^\textbf{x}$ (central mode of back-scattered field of the 1-soliton state), respectively. Moreover, the lower-right panel of Figure~\ref{fig:time-integration} shows the long-term stabilization of the laser field as in \eqref{eq:laser_frequ_convergence}, where the frequency detuning is automatically adjusted to sustain a soliton solution.

\medskip

\begin{figure*}
    \centering
    \includegraphics[width=.37\textwidth]{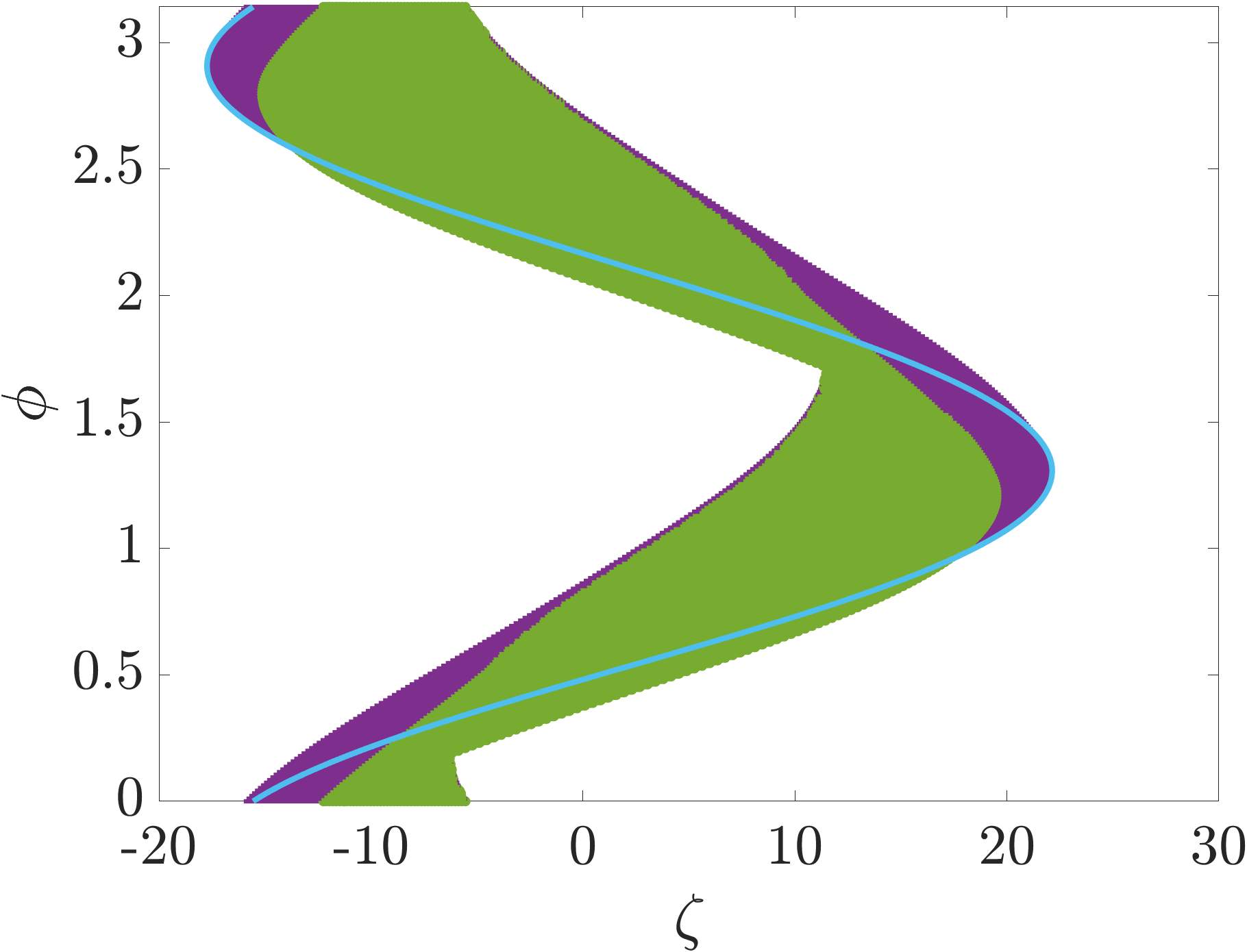}\hspace{4em}
    \includegraphics[width=.365\textwidth]{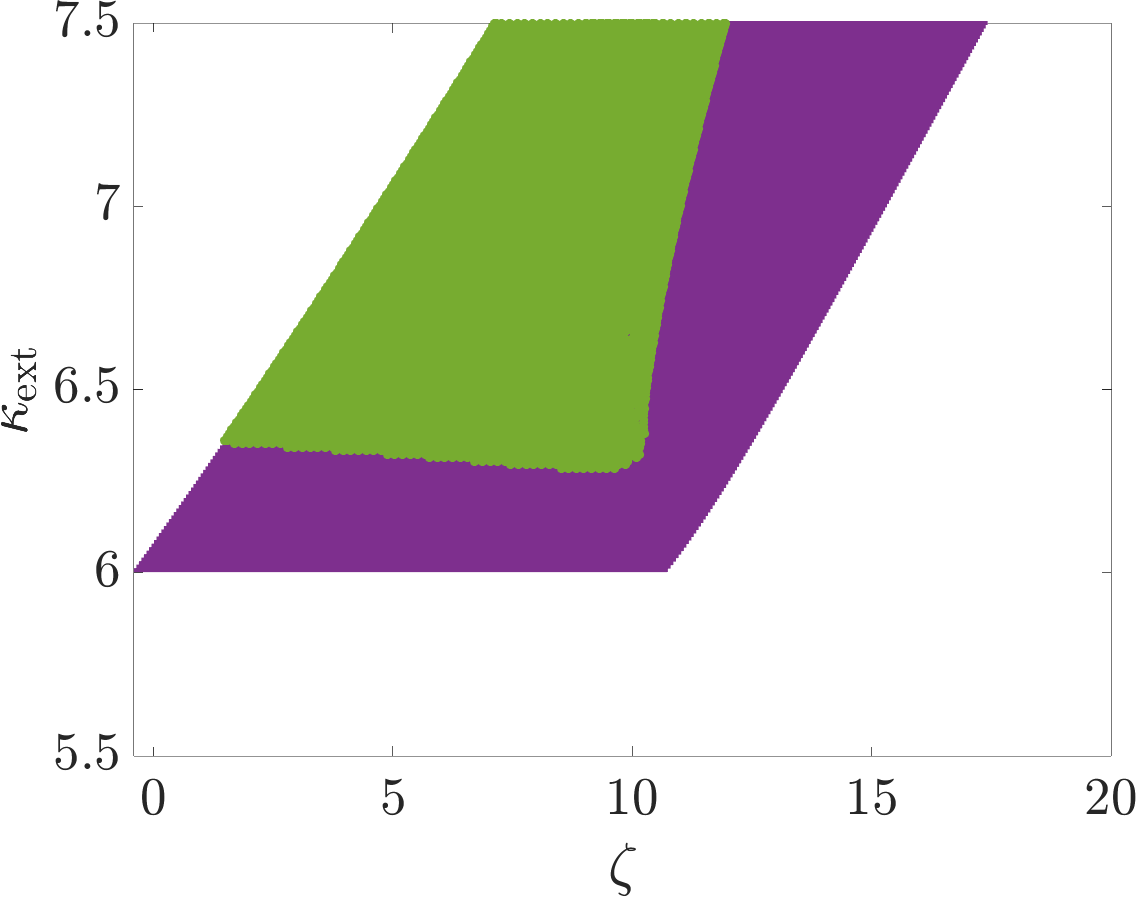}
    \caption{Left: Existence chart in the $\zeta$–$\phi$ plane of stationary 1-soliton solutions to~\eqref{eq:system_in_polar}, bifurcating from flat homogeneous states. Colored area (purple) shows the 1-soliton existence range. The blue curve indicates the position of the BP from which these states emerge. Right: Existence range (purple) in the $\zeta$–$\kappa_\text{ext}$ plane. In both figures, the green subsets depicts the range of 1-solitons with FWHM of at most $0.4$. The other parameters are fixed to the physical values in Table~\ref{tab:constants} with $\kappa_\text{ext} = 6.455$ in the left panel and $\phi = 0.64\times\pi$ in the right panel.}
    \label{fig:existence_chart}
\end{figure*}

Next, we systematically analyze the influence of the parameters $\phi$ and $\kappa_\text{ext}$ on the existence range of 1-soliton solutions. This is of particular interest for practical applications as both parameters can be changed in the experimental set-up~\cite{Shen2020Integrated,Xiang2021Laser,ulanov2024synthetic}. The sensitivity analysis relies on an algorithm based on numerical bifurcation theory, which is explained in Appendix~\ref{app:numerical_bifurcation}. The results are shown in Figure~\ref{fig:existence_chart}. Here, each point in the colored areas represents a pair $(\zeta,\phi)$ (left panel) or $(\zeta,\kappa_\text{ext})$ (right panel) for which 1-soliton can be found. The green areas indicate subsets of the existence range which consist of highly localized 1-soliton with a prescribed full width at half maximum (FWHM) value of at most $0.4$. In addition to that, the blue line in the left panel shows the values of the parameters at which these solitons bifurcate from the homogeneous steady states.

For the accumulated phase $\phi$, we observe that the existence range bends from negative values of $\zeta$ to positive values, and then back to negative, as $\phi$ increases from $0$ to $\pi$. We note that due to the symmetry
$$
    (\un{a}_\text{F},\un{a}_\text{B},\un{a}_\text{L},n,\phi) \mapsto (\un{a}_\text{F},\un{a}_\text{B},-\un{a}_\text{L},n,\phi+\pi)
$$
the existence range is $\pi$-periodic in $\phi$ and thus we can deduce from Figure~\ref{fig:existence_chart} the existence range for all values of $\phi\in [0,2\pi]$. The right panel of Figure~\ref{fig:existence_chart} shows the sensitivity with respect to the coupling $\kappa_\text{ext}$ where the remaining parameters are fixed to the physical values in Table~\ref{tab:constants}. We see that the minimal value of $\kappa_\text{ext} \approx 6.0$ is needed in order to form bifurcating branches supporting 1-soliton solutions. As $\kappa_\text{ext}$ increases beyond a second value, we find localized solitons with a FWHM of at most $0.4$. This is explained by the fact that low values of $\kappa_\text{ext}$ lead to a small excitation of the cavity field obstructing the existence of highly localized states, whereas high values allow energy transfer from the laser to the microresonator resulting in strong forcing. The observation that a strong feedback power is required for 1-soliton formation is in agreement with the experiments in~\cite{ulanov2024synthetic}.

\section{Summary}\label{sec:Summary}
We have investigated the bi-directionally coupled system~\eqref{eq:coupled_system} of a cw-laser and a Kerr-nonlinear microresonator where the backward propagating resonator field interacts with the pump laser. Using numerical bifurcation tools we have shown that the system supports stable time-harmonic $1$-soliton states with stationary normalized instantaneous laser frequency $\omega_\text{L} = \omega_0+\Delta\zeta$. These states have enhanced stability properties in the sense that under sufficiently small perturbations of the unknowns not only the laser and microresonator fields relax towards the $1$-solitons but most importantly $\omega_\text{L}(t)$ self-corrects towards the value $\omega_\text{L}$ which sustains the $1$-soliton state. Such a dynamic self-correction of the laser is impossible for the standard LLE where the laser field has a static frequency. Our approach also offers the advantage of systematically exploring existence ranges of $1$-soliton with respect to parameter variations. We have investigated the sensitivity with respect to accumulated phase $\phi$ and external coupling strength $\kappa_\text{ext}$. It turns out that $1$-solitons can only exist provided $\phi$ lies within a finite band of values and the external coupling strength $\kappa_\text{ext}$ exceeds a minimal value. 

\section*{Acknowledgements}
This project is funded by the Deutsche Forschungsgemeinschaft (DFG, German Research Foundation) -- Project-ID 258734477 -- SFB 1173. The authors thank Tobias Jahnke for providing the MATLAB code used for the time-integration simulations.

\appendix

\section{Numerical bifurcation and sensitivity analysis}\label{app:numerical_bifurcation}
Here we explain the set-up of the numerical bifurcation analysis and we introduce the algorithm which we use for analyzing the sensitivity of the 1-soliton existence range with respect to parameter variations.

\medskip

First note that if the unknowns $a_\text{L}, a_\text{F}$ and $a_\text{B}$ in~\eqref{eq:coupled_system} are all shifted by a common constant phase factor then they still solve the same system. This means that in contrast to the standard Lugiato-Lefever equation~\eqref{eq:lle}, which is only invariant under spatial shifts, the system~\eqref{eq:coupled_system} has an additional invariance under phase shifts of $a_\text{L}, a_\text{F}, a_\text{B}$. We eliminate the phase invariance by transitioning into a rotating frame in which the laser field $a_\text{L}$ is real-valued. This is necessary for a robust convergence of the numerical bifurcation algorithm as it excludes a continuation of nontrivial states in the direction where all components are shifted by a constant phase factor. To this end, we write the laser field in polar coordinates
$$
a_\text{L}(t) = r(t)\eu^{-\iu \int_0^t (\omega_\text{L}(s)-\omega_\text{m}) \de s}, \quad r(t)=|a_\text{L}(t)|,
$$
where we recall that $\omega_\text{L}(t)-\omega_\text{m} = -\frac{d}{dt} \arg(a_\text{L}(t))$. Moreover, we introduce $\un{a}_\text{F}, \un{a}_\text{B}$ defined by
\begin{align*}
a_\text{F}(x,t) & = \un{a}_\text{F}(x,t)\eu^{-\iu \int_0^t (\omega_\text{L}(s)-\omega_\text{m}) \de s}, \\ 
a_\text{B}(t) & = \un{a}_\text{B}(t)\eu^{-\iu \int_0^t (\omega_\text{L}(s)-\omega_\text{m}) \de s}.
\end{align*}
Substituting the formula for $a_\text{L}$ into~\eqref{eq:coupled_system} we obtain an equation for the instantaneous laser frequency 
\begin{align} \label{eq:laser_frequency}
\begin{split}
\omega_\text{L}(t) =& \omega_0 +\frac{1}{2} \alpha_\text{H}
    \left(g(r(t)^2) n_0(n(t)-1)- \kappa_\text{d}\right) \\
    & + \frac{\kappa_\text{ext}}{r(t)} \Im \left(\un{a}_\text{B}(t) \eu^{\iu \phi}\right).
    \end{split}
\end{align}
With the shorthand 
$$
\Theta(r,z,n) :=  \frac{1}{2} \alpha_\text{H}
    \left(g(r^2)n_0 (n-1)- \kappa_\text{d}\right) +\frac{\kappa_\text{ext}}{r} \Im \left(z \eu^{\iu\phi} \right)
$$
and the frequency offset $\zeta=\omega_\text{m}-\omega_0$ the remaining system for $n, r, \un{a}_\text{F}, \un{a}_\text{B}$ becomes (we suppress their arguments):
\begin{align}\label{eq:system_in_polar}
\begin{split}
    \dot n =& \iota - \gamma n - g(r^2) (n-1) r^2, \allowbreak \\
    \dot r =& 
    \frac{1}{2}\left(g(r^2)n_0 (n-1)- \kappa_\text{d}\right)r - \kappa_\text{ext}\Re\left( \un{a}_\text{B} \eu^{\iu \phi}\right),\allowbreak \\
    \dot{\un{a}}_\text{F} =& \Bigl[ - 1- \iu \left(\zeta-\Theta(r,\un{a}_\text{B},n)\right) + \iu d_2 \frac{\partial^2}{\partial x^2} \\
    & + \iu  (|\un{a}_\text{F}|^2 + 2 |\un{a}_\text{B}|^2) \Bigr] \un{a}_\text{F} + \iu \kappa_\text{sc}\un{a}_\text{B} - \kappa_\text{ext} r \eu^{\iu \phi}, \allowbreak \\
    \dot{\un{a}}_\text{B} =& \Biggl[ \left(-1 - \iu (\zeta-\Theta(r,\un{a}_\text{B},n)\right) \allowbreak\\
    & + \iu\left( |\un{a}_\text{B}|^2 + \frac{2}{2\pi} \int_0^{2\pi} |\un{a}_\text{F}(x,\cdot)|^2 \de x \right) \Biggr] a_\text{B} \\
    & + \iu \overline{\kappa}_\text{sc} \frac{1}{2\pi} \int_0^{2\pi} \un{a}_\text{F}(x,\cdot) \de x.
\end{split}
\end{align}
Observe that in this system the invariance under phase-shifts is no longer present since $r$ is a real-valued quantity.

When searching for steady states of~\eqref{eq:system_in_polar} we set the time-derivatives $\dot n,\dot r,\dot{\un{a}}_\text{F},\dot{\un{a}}_\text{B} $ to zero. This results in a ODE for $\un{a}_\text{F}$ coupled to three algebraic equations for the remaining fields which include nonlocal terms of $\un{a}_\text{F}$. For the instantaneous laser frequency $\omega_\text{L}$, equation~\eqref{eq:laser_frequency} implies that it will also be constant. This means that stationary solutions of~\eqref{eq:system_in_polar} correspond to time-harmonic solutions of~\eqref{eq:coupled_system}.

Combing~\eqref{eq:laser_frequency} and the equation for $r$ in~\eqref{eq:system_in_polar} we obtain
\begin{align}\label{eq:laser_frequency_steady}
    \omega_\text{L} = \omega_0 + \frac{\kappa_\text{ext}}{r} \Im\left(\un{a}_\text{B} \eu^{\iu \phi}(1-\iu \alpha_\text{H})\right).
\end{align}
The steady equation for $n$ can be solved in terms of $r$, yielding the formula
$$
    n = n(r) = \frac{\iota + g(r^2)r^2}{\gamma + g(r^2)r^2}.
$$
Upon substituting this into the system~\eqref{eq:system_in_polar}, we obtain an ODE for $\un{a}_\text{F}$, coupled to two algebraic equations for $r$ and $\un{a}_\text{B}$, given by
\begin{align}\label{eq:system_reduced}
\begin{split}
    0 =& \Bigl[ - 1- \iu \left(\zeta-\Theta(r,\un{a}_\text{B},n(r))\right) + \iu d_2 \frac{\partial^2}{\partial x^2} \\
    & + \iu  (|\un{a}_\text{F}|^2 + 2 |\un{a}_\text{B}|^2) \Bigr] \un{a}_\text{F} + \iu \kappa_\text{sc}\un{a}_\text{B} - \kappa_\text{ext} r \eu^{\iu \phi}, \\
    0 =& 
    \frac{1}{2}\left(g(r^2)n_0 (n(r)-1)- \kappa_\text{d}\right)r - \kappa_\text{ext}\Re\left( \un{a}_\text{B} \eu^{\iu \phi}\right),\\
    0 =& \Biggl[ \left(-1 - \iu (\zeta-\Theta(r,\un{a}_\text{B},n(r))\right) \\
    & + \iu\left( |\un{a}_\text{B}|^2 + \frac{2}{2\pi} \int_0^{2\pi} |\un{a}_\text{F}(x)|^2 \de x \right) \Biggr] \un{a}_\text{B} \\
    & + \iu \overline{\kappa}_\text{sc} \frac{1}{2\pi} \int_0^{2\pi} \un{a}_\text{F}(x) \de x.    
\end{split}
\end{align}
We consider the ODE for $\un{a}_\text{F}$ on the interval $x\in [0,\pi]$ equipped with homogeneous Neumann boundary conditions
\begin{align}\label{eq:Neumann_bc}
    \partial_x\un{a}_\text{F}(0) = \partial_x\un{a}_\text{F}(\pi) = 0.
\end{align}
This eliminates the spatial shift invariance of $\un{a}_\text{F}$ in~\eqref{eq:system_in_polar}, which is necessary to avoid unwanted path continuation in the translational direction. Note that we recover periodic solutions on $[0, 2\pi]$ by reflection: 
$$
    \un{a}_{\text{F},\text{per}}(x)=
    \begin{cases}
        \un{a}_\text{F}(x), & x \in [0,\pi],\\
        \un{a}_\text{F}(2\pi-x), & x \in [\pi,2\pi].
    \end{cases}
$$
We then perform the numerical bifurcation analysis for~\eqref{eq:system_reduced}-\eqref{eq:Neumann_bc} . The bifurcation analysis is conducted using the MATLAB package \texttt{pde2path}~\cite{Uecker2014pde2path}, which is designed for the numerical continuation and bifurcation analysis of systems of PDEs. We discretize the boundary value problem for $\un{a}_\text{F}$ using linear finite elements, and add the equations for $\un{a}_\text{B}$ and $r$ as algebraic constraints. Throughout our bifurcation experiments, we choose the frequency offset $\zeta$ as the bifurcation parameter, since it can be varied in the physical experiments.

\medskip

\begin{figure}
    \centering
    \includegraphics[width=.35\textwidth]{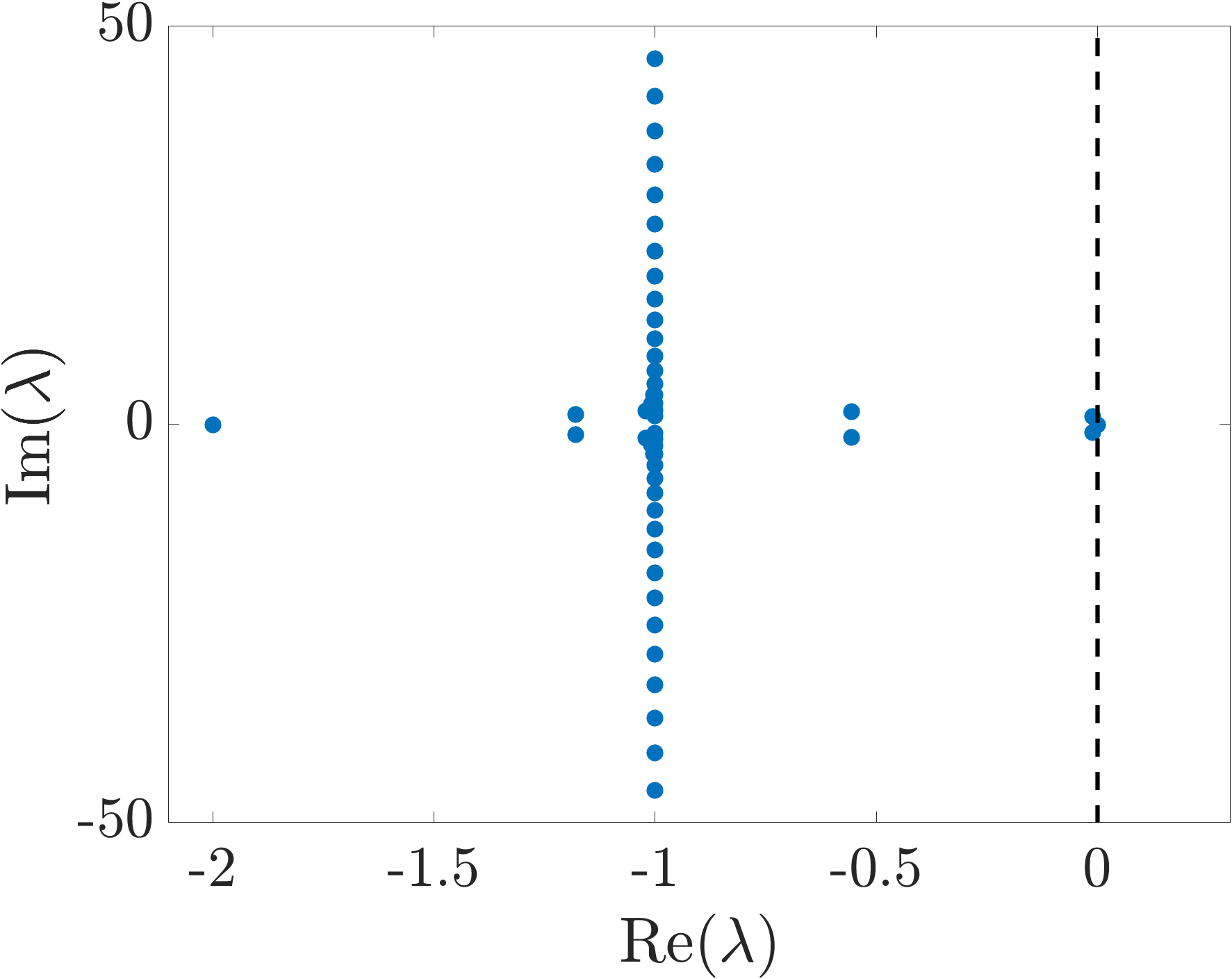}
    \caption{Eigenvalues of the Jacobian operator evaluated at the 1-soliton state with label \textbf{x} in Figure~\ref{fig:bif_run1}. We see that all eigenvalues except for the zero eigenvalue have strictly negative real part indicating dynamical stability and relaxation to a spatial translate of the 1-soliton solution at an exponential rate.}
    \label{fig:soliton_spectrum}
\end{figure}

Stability of the bifurcating solutions can be determined by computing the eigenvalues of the Jacobian $\mathcal{J}$ associated with the discretized version of~\eqref{eq:system_reduced}. For this purpose it is necessary to reintroduce the translational invariance by using periodic boundary conditions on $[0,2\pi]$ for the field $\un{a}_\text{F}$ as otherwise only stability with respect to even perturbations is determined. Thus, the solution is stable if all eigenvalues $\lambda$ of $\mathcal{J}$ satisfy $\Re(\lambda)< 0$ with the exception of a zero eigenvalue present due to the continuous shift symmetry of the system. As a result, stable solutions exhibit a spectral gap (if we neglect to zero eigenvalue), which causes a relaxation of the perturbed solutions to a spatial translate of it on exponentially fast timescales, cf.~Figure~\ref{fig:time-integration}. Figure~\ref{fig:soliton_spectrum} shows the spectrum of the 1-soliton states $\textbf{x}$ indicating its stability.

\medskip

The bifurcation analysis is also used to algorithmically determine (a subset of) the $1$-soliton existence range in the parameter space. To describe our algorithm for the sensitivity analysis, we need an auxiliary result from bifurcation theory. Suppose that $(\un{a}_{\text{F},0},r,\un{a}_{\text{B},0},\zeta_0) \in \C \times \R \times \C\times \R$ is a nondegenerate BP on the constant solution curve, where nondegenerate means that the associated Jacobian of~\eqref{eq:system_reduced} has a simple zero eigenvalue. Then the forward field component of the bifurcating solutions can be locally parametrized by $s\in (-s_0,s_0)$ with some $0<s_0\ll 1$,~\cite[Corollary I.4.2]{Kielhoefer2012Bifurcation}. More precisely, we have the expansion
\begin{align}\label{eq:asysmptotic_bp}
    \un{a}_\text{F}(x;s) = \un{a}_{\text{F},0} +\un{a}_{\text{F},1} \cos(kx)s + \mathcal{O}(s^2) \text{ as }s \to 0
\end{align}
with $k \in \N_0$ and $\un{a}_{\text{F},1} \in \C$. Below, we will use this formula for $k=1$ to determine those BPs at which 1-solitons branch away from flat states. Now let us fix a system parameter $\varsigma \not = \zeta$. We analyze the influence of $\varsigma$ on the existence range of the 1-solitons. Therefore, we chose a discretization $\varsigma_\text{min}=\varsigma_1<\varsigma_2<\dots<\varsigma_m=\varsigma_\text{max}$ with $m\in \N$. Our algorithm consists of three steps.

\medskip

\noindent\textbf{Algorithm:} For $j=1,\dots,m$ do:
\begin{itemize}
    \item[1.] Set $\varsigma = \varsigma_j$ and compute the constant solution curve with bifurcation parameter $\zeta \in [\zeta_{\min},\zeta_{\max}]$.
    \item[2.] Mark all nondegenerate BPs on this curve for which~\eqref{eq:asysmptotic_bp} holds with $k=1$.
    \item[3.] Compute and plot all bifurcation branches starting from the marked BPs of Step 2 in $\zeta$ as long as the forward field component $\un{a}_\text{F}$ satisfies the following conditions:
    $$
    \begin{array}{lc}
    \text{(i)}     &  \{x_{\max}\} =\argmax_x|\un{a}_\text{F}(x)|^2 \subset \{0,\pi\}, \\
    \text{(ii)} & \left|\frac{\int_{\pi/2}^{x_\text{max}} |\un{a}_\text{F}'(z)|^2 \de z }{\int_{\pi/2}^{\pi-x_{\text{max}}} |\un{a}_\text{F}'(z)|^2 \de z }\right|> 1.75. 
    \end{array}
    $$
\end{itemize}
We briefly comment on each step. Steps 1 and 2 are necessary to compute nontrivial solutions of~\eqref{eq:system_reduced} where Step~2 ensures that we only calculate bifurcation branches on which we expect to find 1-soliton states. Heuristically, we have observed that $k_0$-solitons emerge from BPs where~\eqref{eq:asysmptotic_bp} holds with $k=k_0$. This observation is consistent with simulations in~\cite{Gaertner2019Bandwidth} for the standard LLE~\eqref{eq:lle}. Once we are on a 1-soliton branch, we stop the continuation as soon as the solutions deform into 2-solitons states (or different wave forms). In system~\eqref{eq:system_reduced} we have observed that a 1-soliton branch generically connects to a 2-soliton branch. The stopping criteria are given by conditions (i) and (ii) in Step~3. The first ensures that the modulus squared of the periodic extension of $\un{a}_\text{F}(x)$ has a single maximum on the periodicity cell $[0,2\pi]$, while the second ensures that the solutions are sufficiently localized around this maximum. Note that the value $1.75$ determines the degree of localization; it can also be varied, with low values meaning that we look for a low degree of localization, while high values enforce a high degree of localization.

\section{Derivation of the normalized system from the physical model}\label{app:normalization}
Here we show how the normalized coupled system~\eqref{eq:coupled_system} is derived from the system in physical units and we give the values of the physical and normalized parameters used in our simulations.

The bi-directionally coupled laser-resonator system in physical units is given by~\cite{Xiang2021Laser,Lihachev2022Platicon}
\begin{align}\label{eq:coupled_system_Physical}
    \begin{split}
    \dot N(t) =& I - \Gamma N(t) - \frac{\mu^2 \sigma}{1+\epsilon |A_\text{L}(t)|^2} (N(t)-N_0) |A_\text{L}(t)|^2, \\
    \dot A_\text{L}(t) =& 
    \Bigl[\frac{1-\iu\alpha_\text{H}}{2} \left(\frac{\mu^2 \sigma}{1+\epsilon |A_\text{L}(t)|^2} (N(t)-N_0)- K_\text{d}\right)\\
    & + \iu (\Omega_\text{m}-\Omega_0)\Bigr]A_\text{L}(t)- K_\text{ext} A_\text{B}(t)\eu^{\iu \phi} \\
    \dot A_\text{F}(x,t) =& \Bigl[ - \frac{K}{2} + \iu D_2 \frac{\partial^2}{\partial x^2} + \iu  g_0(|A_\text{F}(x,t)|^2 + 2 |A_\text{B}(t)|^2)\Bigr] \\
    & \times A_\text{F}(x,t) + \iu K_\text{sc}A_\text{B}(t) - K_\text{ext} A_\text{L}(t) \eu^{\iu \phi}, \\
    \dot A_\text{B}(t) =& \Biggl[ -\frac{K}{2} + \iu g_0 \bigl( |A_\text{B}(t)|^2 + \frac{1}{\pi} \int_0^{2\pi} |A_\text{F}(x,t)|^2 \de x \bigr)\Biggr] \\
    & \times A_\text{B}(t) + \iu \overline{K}_\text{sc} \frac{1}{2\pi} \int_0^{2\pi} A_\text{F}(x,t) \de x.
    \end{split}
\end{align}
The physical meanings of the constants are explained in Table~\ref{constants_table} and the normalized constants are given by
\begin{align*}
    \iota &= 2\frac{I}{KN_0}, \quad \gamma=2\frac{\Gamma}{K},\quad
    \kappa_\text{d} = 2\frac{K_\text{d}}{K}, \\
    \omega_\text{m}&= 2 \frac{\Omega_\text{m}}{K},\quad \omega_0= 2 \frac{\Omega_0}{K}, \quad 
    \kappa_\text{ext} = 2 \frac{K_\text{ext}}{K}, \quad
    d_2 = 2\frac{D_2}{K}, \\
    \kappa_\text{sc} &= 2\frac{K_\text{sc}}{K},\quad
    n_0 = 2 \frac{N_0}{K \mu}, \quad
    \eps = \frac{\epsilon K}{2}.
\end{align*}
The system~\eqref{eq:coupled_system} is then obtained by the transformation
\begin{align*}
    a_\text{F}(x,t) &= \sqrt{\frac{2 g_0}{K}} A_\text{F}\left(x,\frac{2}{K}t\right), \\
    a_\text{B}(t) & = \sqrt{\frac{2 g_0}{K}} A_\text{B}\left(\frac{2}{K}t\right), \\
    a_\text{L}(t) &= \sqrt{\frac{2 g_0}{K}} A_\text{L}\left(\frac{2}{K}t\right), \\
    n(t) &= N_0^{-1} N\left(\frac{2}{K}t\right).\\
    & \\
    & \\
    & \\
    & \\
 \end{align*}

\begin{table}
\begin{tabular}{|c|c|c|c|} \hline
quantity & value & normal. & value \\ 
& & quantity & \\ \hline
$I/e$  & $\frac{0.5}{1.6\times 10^{-19}}$ A/C & $\iota$ & $41.4466$ \\
$\Gamma$ & $1\times 10^{9}$ s$^{-1}$  & $\gamma$ & $2.6526$\\
$N_0$ & $2\times 10^8$ & $n_0$ & $1.0610$\\
$\mu$ &  $0.5$ & $-$ & $-$\\
$\alpha_\text{H}$ & $2.5$ & $-$ & $-$\\
$K_\text{d}$ & $2\times 10^{10}\times 2\pi$ rad s$^{-1}$ & $\kappa_\text{d}$  & $833.3333$\\
$K_\text{ext}$ & $[5.5,7.5]\times 120\times10^6\times \pi$ rad s$^{-1}$ & $\kappa_\text{ext}$ & $[5.5,7.5]$\\
$K$ & $120 \times 10^6 \times 2\pi$ rad s$^{-1}$ & $-$ & $-$\\
$D_2$ & $7.5 \times 10^6 \times 2 \pi$ rad s$^{-1}$ & $d_2$ & $0.125$\\
$g_0$ & $1.866$ rad$^{-1}$ & $-$ & $-$\\
$K_\text{sc}$ & $15\times 10^6 \times 2\pi $ rad s$^{-1}$ & $\kappa_\text{sc}$ & $0.25$\\ 
$\phi$ & $[0,\pi]$ & $-$ & $-$ \\
$\sigma$ & $1 \times 10^4$ s$^{-1}$ & $-$ & $-$\\
$\epsilon$ & $0$ & $\eps$ & $0$\\
\hline
\end{tabular}
\caption{Values of the physical constants and their normalized counterparts. A dash indicates that there is no normalized version of the corresponding quantity in the equations.}
\label{tab:constants}
\end{table}

\bibliographystyle{apsrev4-1}
\bibliography{bibliography}

\end{document}